\documentclass{PoS}

\def\gsim{\mathrel{\rlap{\lower4pt\hbox{\hskip1pt$\sim$}}
    \raise1pt\hbox{$>$}}}         
\def\lsim{\mathrel{\rlap{\lower4pt\hbox{\hskip1pt$\sim$}}
    \raise1pt\hbox{$<$}}}         

\newcommand{\be}{\begin{equation}}
\newcommand{\ee}{\end{equation}}
\newcommand{\bea}{\begin{eqnarray}}
\newcommand{\eea}{\end{eqnarray}}
\newcommand{\bi}{\begin{itemize}}
\newcommand{\ei}{\end{itemize}}
\newcommand{\ben}{\begin{enumerate}}
\newcommand{\een}{\end{enumerate}}

\newcommand{\lp}{\left(}
\newcommand{\rp}{\right)}

\title{Machine Learning tools for global PDF fits}

\ShortTitle{Machine Learning tools for global PDF fits}

\author{\speaker{Juan Rojo}\\
  Department of Physics and Astronomy,
VU University, De Boelelaan 1081,
1081 HV Amsterdam,\\ 
and Nikhef Theory Group,
Science Park 105, 1098 XG Amsterdam,
The Netherlands. \\
        E-mail: \email{j.rojo@vu.nl}}

\abstract{
  The use of machine learning algorithms in theoretical
  and experimental high-energy
  physics has experienced an impressive progress in recent
  years, with applications from trigger selection to jet
  substructure classification and detector simulation among many others.
  In this contribution, we review the machine learning tools used
  in the NNPDF family of global QCD analyses.
  These include multi-layer feed-forward neural networks
  for the model-independent parametrisation of
  parton distributions
  and fragmentation functions, genetic and covariance matrix adaptation
  algorithms for training and optimisation, and closure testing
  for the systematic validation of the fitting methodology.
}

\FullConference{XXIIIth Quark Confinement and the Hadron Spectrum\\
		1-6 August 2018\\
		University of Maynooth, Ireland}

\begin{document}

\vspace{0.1cm}
\noindent

\paragraph{Machine learning and high energy physics.}

The recent years have experienced an unprecedented boost
in the quantity and quality of the applications of machine learning (ML)
algorithms in both theoretical
and experimental high-energy physics, as summarised in the recent
Community White Paper~\cite{Albertsson:2018maf} and in the
review~\cite{Radovic:2018dip}.
Similar developments have taken place in related areas such as astroparticle
physics, the intensity frontier, and cosmology.
In the context of LHC physics, machine learning tools
have been exploited in applications ranging from detector
simulation~\cite{Paganini:2017dwg} to the exploration of the
parameter space of New Physics scenarios~\cite{Caron:2016hib}
and the study of the substructure of hadronic jets~\cite{Larkoski:2017jix},
among several others.
ML tools are even used in formal theory studies,
for instance to explore the huge number of possible
solutions (the landscape) predicted by string theory~\cite{Carifio:2017bov}.

In the specific case of theoretical high-energy physics,
a possible classification of the ML algorithms used was proposed
in~\cite{Carrazza:2017qro}.
In the first category, one finds computational techniques and tools
relevant for advanced numerical methods,  Monte Carlo event generators,
and higher-order perturbative calculations and computer algebra
among others.
The second category includes applications of modern ML
techniques such as supervised learning (regression and
classification), uncertainty propagation, and even
``experimental'' mathematics.

\begin{figure}[t]
  \centering
  \includegraphics[width=.45\linewidth,angle=-90]{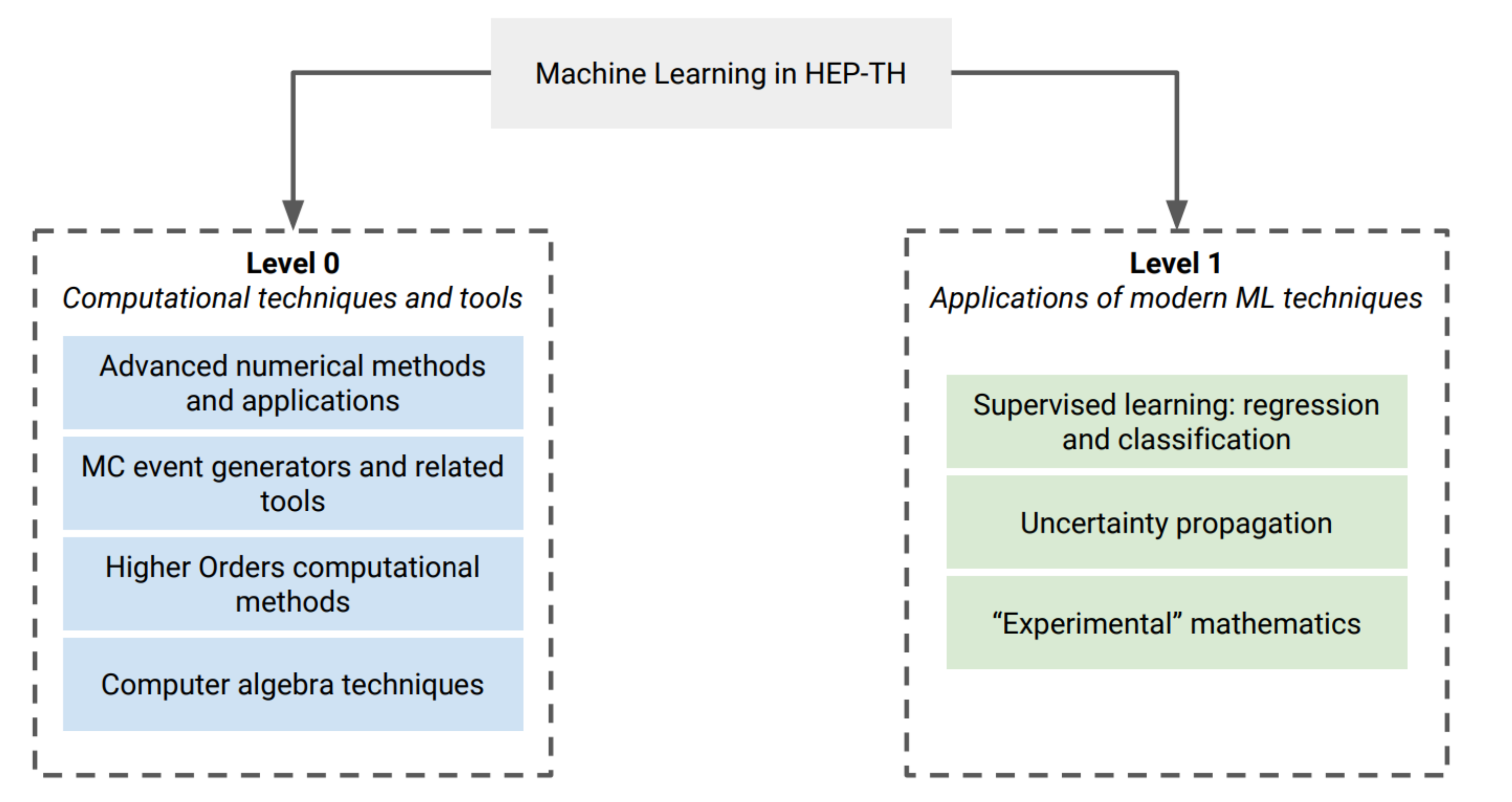}
  \caption{\small Machine learning tools in theoretical
    high-energy physics can be divided into two
    main categories: {\it (i)} computational
    techniques and tools,
    and {\it (ii)} applications
    of modern ML methods.
    Figure taken from~\cite{Carrazza:2017qro}.
  }
\label{fig:summary}
\end{figure}

In the context of LHC phenomenology, an
important ingredient of theoretical predictions
for event rates are the parton
distribution functions (PDFs) of the proton~\cite{Gao:2017yyd}, describing
the momentum distribution that quarks and
gluons (as well as photons~\cite{Bertone:2017bme})
carry in the initial stage of the hadronic collision.
These PDFs are determined from non-perturbative dynamics, and therefore
their computation from first principles is extremely challenging (see
however encouraging progress from lattice QCD~\cite{Lin:2017snn}).
Therefore, they need to be extracted from experimental
data by means of the so-called global QCD analysis.
As discussed in this contribution,
also for this specific
application machine learning tools have shown to be highly effective,
from the use of artificial neural networks as universal unbiased interpolants
to advanced optimisation strategies for the exploration of complex
high-dimensional parameter spaces.

\begin{figure}[t]
  \centering
  \includegraphics[width=.62\linewidth,angle=-90]{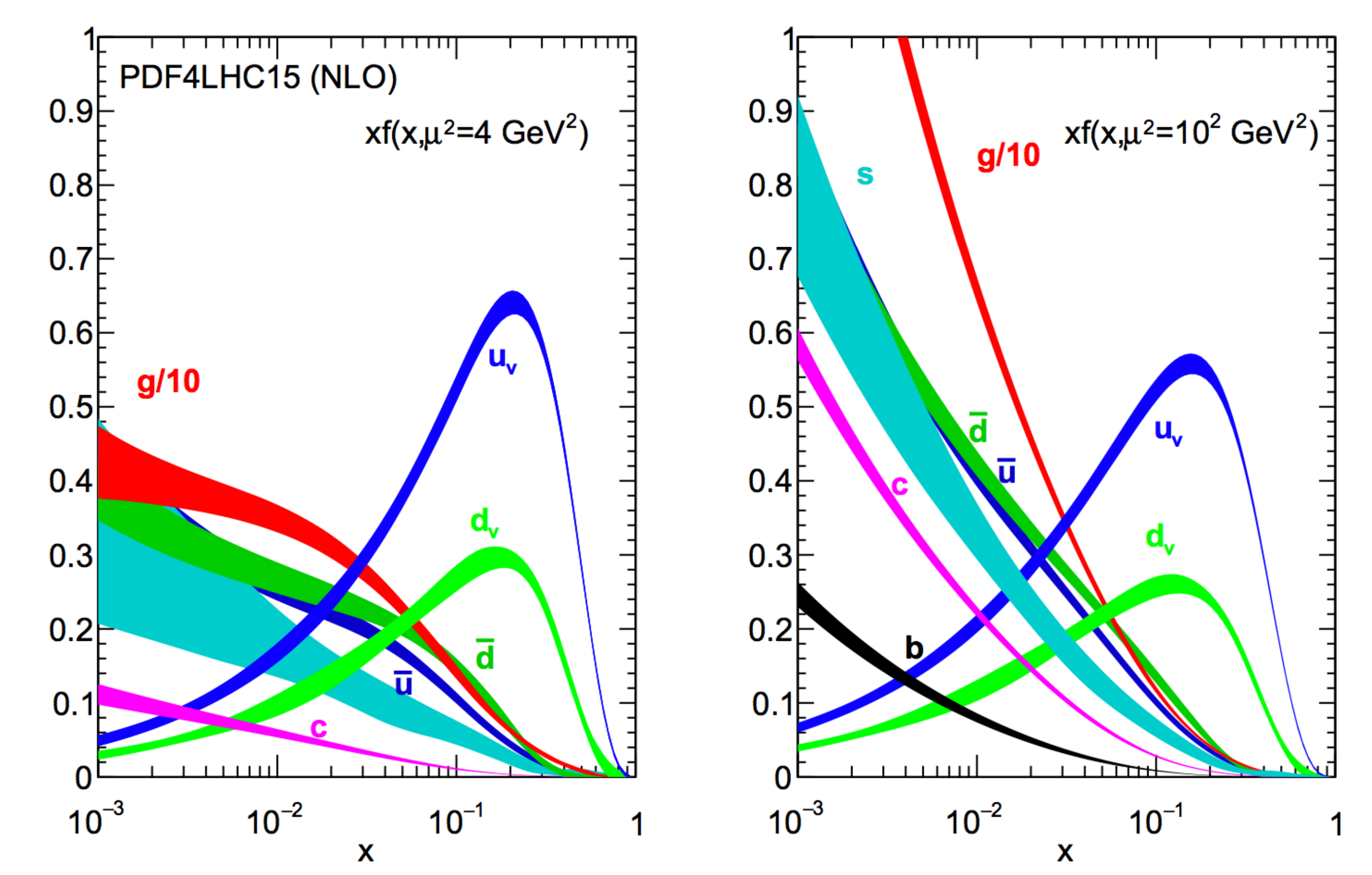}
  \caption{\small The parton
    distributions from
    the PDF4LHC15 NLO set for a low scale, $Q=2$ GeV, where
    the PDFs are parametrised,
    and a higher scale $Q=10$ GeV, highlighting the
    impact of DGLAP evolution.
     }
\label{fig:pdf4lhc}
\end{figure}

In a nutshell, the main task of the global QCD analysis is
to determine the parton distributions of the proton at some
low scale $Q_0$, and then evolve them upwards
to higher energies to carry out predictions
for LHC processes, say for Higgs production.
In Fig.~\ref{fig:pdf4lhc}  we show the NLO PDF4LHC15
set~\cite{Butterworth:2015oua,Carrazza:2015hva}
both at $Q=2$ GeV, close to the typical scale where the
PDFs are parametrised,
and at a higher scale $Q=10$ GeV, highlighting the
effects of DGLAP evolution.
We can observe how for $x\lsim 0.01$ the perturbative
QCD evolution drives a steep rise in the gluon
and the sea quarks.

\paragraph{Machine learning tools in the NNPDF framework.}

The NNPDF approach to global QCD analyses has been successfully
applied to the determination of the unpolarised~\cite{Ball:2017nwa}
and polarised~\cite{Nocera:2014gqa} parton distributions of the proton
as well as to the light-hadron identified and the unidentified
fragmentation functions (FFs)~\cite{Bertone:2017tyb,
  Bertone:2018ecm}.
The latter are the time-like counterparts
of the PDFs, and describe the hadronisation process of colored partons
into color neutral-hadrons.
There is also work in progress towards NNPDF fits of nuclear
modification factors, relevant for the interpretation of heavy-ion
collisions at RHIC and the LHC.

\begin{figure}[t]
  \centering
  \includegraphics[width=.99\linewidth]{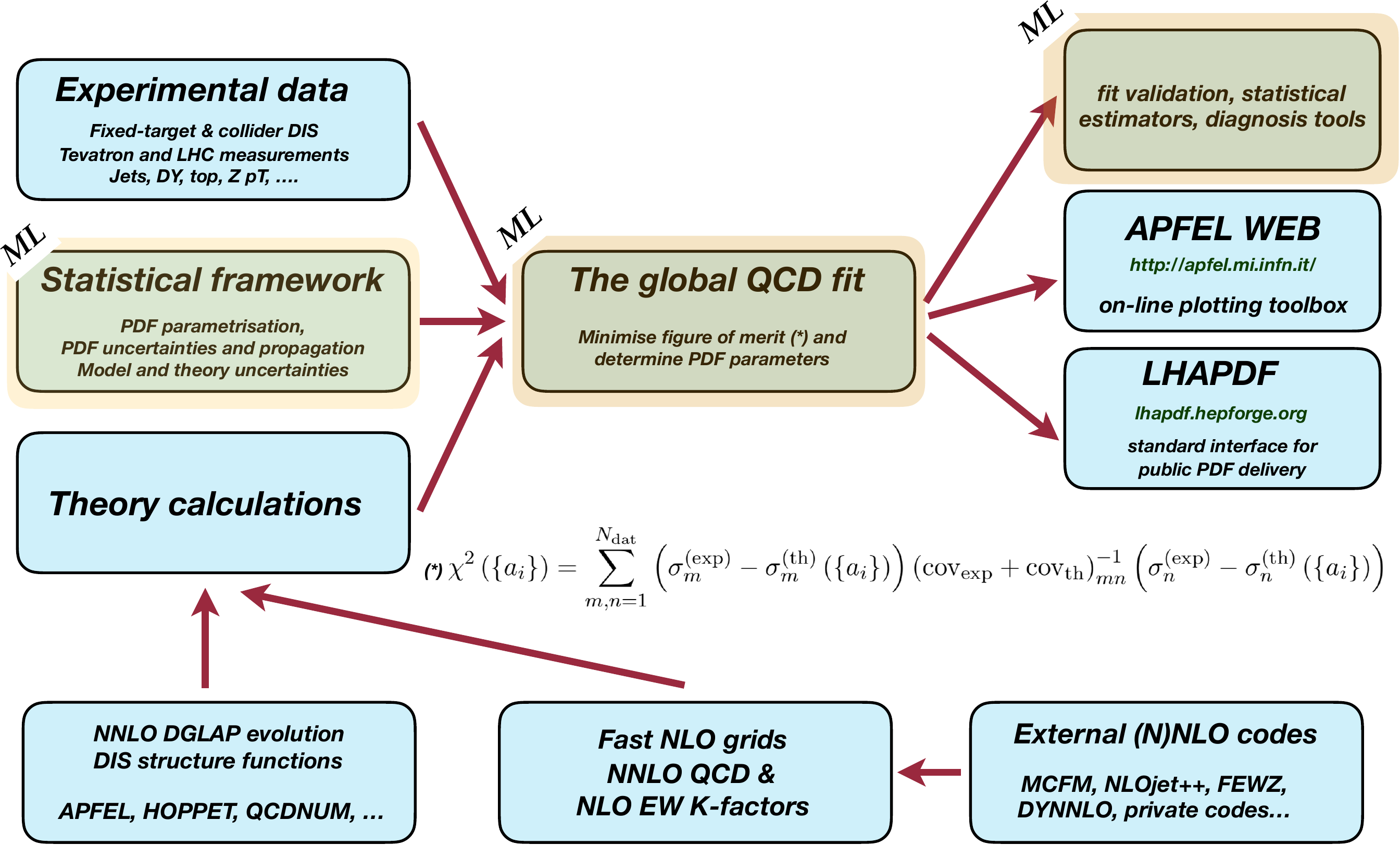}
  \caption{\small Representation of the different components
    that constitute the NNPDF family of global QCD analysis,
    highlighting those that exploit machine learning tools.
  }
\label{fig:PDFfit}
\end{figure}

In Fig.~\ref{fig:PDFfit} we indicate the different components
    that constitute the NNPDF family of global analyses,
    highlighting those that involve machine learning algorithms.
    As illustrated there,
    a global QCD fit is based on three main inputs: experimental
    data, higher-order perturbative calculations in both QCD
    and QED/electroweak theory, and a statistical framework
    dealing with aspects such as
    the PDF parametrisation and their uncertainty estimate and propagation.
    
    These three ingredients are combined in the global QCD fit
    by means of the minimisation of a suitably defined figure of merit, the $\chi^2$,
    which includes all relevant sources of uncertainty and which leads 
    to the determination of the parameters that define the PDF shape.
    Experimental uncertainties are propagated to the PDFs by means
    of the Monte Carlo replica method, which allows constructing a representation
    of the probability density in the space of PDFs.
    Afterwards, the fit is validated using a range of
    complementary diagnosis tools and it can be plotted in
    different ways.
    Finally, the PDF fit
    is translated into the
    {\tt LHAPDF} standard interface, suitable for
    its public delivery and its integration into other HEP codes
    and into the analysis framework of the LHC experiments.
    
    As shown in Fig.~\ref{fig:PDFfit}, machine learnings tools arise in
    various components of the NNPDF  framework, including
    the strategy for the PDF parametrisation,
     the optimisation (training)
     that defines the best-fit parameters, and the subsequent validation
     by means of closure testing.
    We describe now each of these aspects in turn.

\paragraph{PDF parametrisation and neural networks.}
In the NNPDF approach, the parton distribution functions
(or the fragmentation functions) are parameterized at a low scale,
around the boundary between the perturbative and non-perturbative
regimes of QCD, namely
$Q_0 \simeq 1$ GeV (the proton mass).
As opposed to other fitting approaches, where the PDF shape is parametrised
in terms of relatively simple functional forms more or less inspired
in QCD models (see~\cite{Ball:2016spl} for an overview), we use artificial
neural networks (NNs) as unbiased interpolants.
This allows us to avoid the theoretical biases that can be incurred
when specific model functional forms are adopted.
Note here that QCD
provides only very limited guidance about the behaviour of PDFs at the
input parametrisation scale $Q_0$, such as integrability conditions
and the momentum and valence sum rules, and does not provide
any further information on their $x$ dependence at low scales.

\begin{figure}[t]
  \centering
  \includegraphics[width=.84\linewidth]{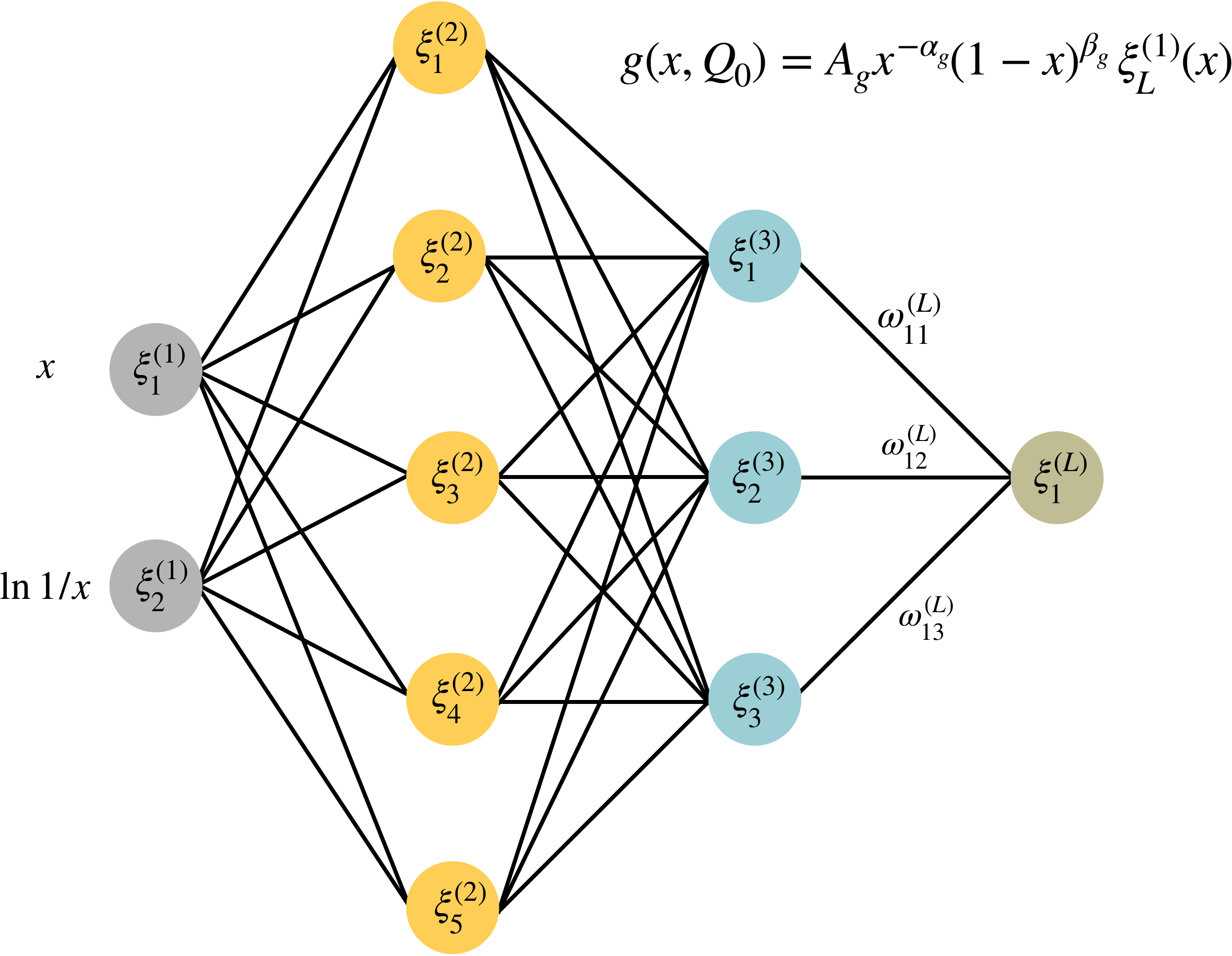}
  \caption{\small A multi-layer feed-forward
    artificial neural network (perceptron) 
    such as the one used in the NNPDF global analysis.
    This specific network has a 2-5-3-1 architecture with two inputs
    ($x$ and  $\ln 1/x$) and one output neuron, which is
    directly related to the value of the PDF at the
    input parametrisation scale $Q_0$, as indicated by Eq.~(\ref{eq:gluonparam}).
  }
\label{fig:ANN}
\end{figure}

Specifically, in the NNPDF fits we use multi-layer feed-forward
artificial neural networks (perceptrons)
such as the one shown in Fig.~\ref{fig:ANN}.
This NN has a  2-5-3-1 architecture with two inputs
    ($x$ and  $\ln 1/x$) and one output neuron, which is
    directly related the the value of the PDF at the
    input parametrisation scale $Q_0$.
    The activation state of each neuron is denoted by $\xi_{i}^{(l)}$, with
    $l$ labelling the layer and $i$ the specific neuron
    within each layer.
    The values of the activation states of the neurons
    in layer $l$ are evaluated in terms of those of the previous
    layer ($l-1$) and the weights $\{\omega_{ij}^{(l)}\}$ connecting
    them as well as by the activation thresholds
    of each neuron $\{\theta_{i}^{(l)}\}$,
    see~\cite{Ball:2008by} and references therein.
    The training of the NN in this context corresponds
    to determining the values of the weights and thresholds
    that fulfill the constraints of a
    given optimisation problem as discussed below.

The value of a given PDF, say the gluon, at the input parametrisation
scale $Q_0$ is then given in terms of the activation state
of the neuron in the last layer as follows
\be
\label{eq:gluonparam}
g(x,Q_0) = A_g x^{-\alpha_g}(1-x)^{\beta_g}\,\xi_1^{(L)}(x) \, ,
\ee
where $A_g$ is here an overall normalisation constant fixed by the momentum
sum rule.
The $x^{-\alpha_g}(1-x)^{\beta_g}$ term is known as the
preprocessing factor and facilitates the NN training by
allowing to learn a smoother function without introducing
any bias in the fits.
The values of the exponents $\{\alpha,\beta\}$ are chosen
at random in an interval determined iteratively~\cite{Ball:2014uwa}.

The use of NNs as universal unbiased interpolants offers a number
of important advantages in the context of the global PDF fit.
In particular, they ensure that the fit results are driven only by the
input data and theory, but not by model-dependent assumptions.
Specifically, we have shown that the NNPDF fits are stable
with respect to changes in the quark flavour basis, the value
of the input scale $Q_0$, and the NN architecture (provided we work
in the redundant regime).
In the latter case, we have verified that even increasing
the number of fitted parameters by an order of magnitude
(from around 400 parameters with
2-5-3-1 to around 4000 parameters with the 2-20-5-1 architecture) leads to
results at the PDF level which are statistically equivalent~\cite{Ball:2014uwa}.

\begin{figure}[t]
  \centering
  \includegraphics[width=.99\linewidth]{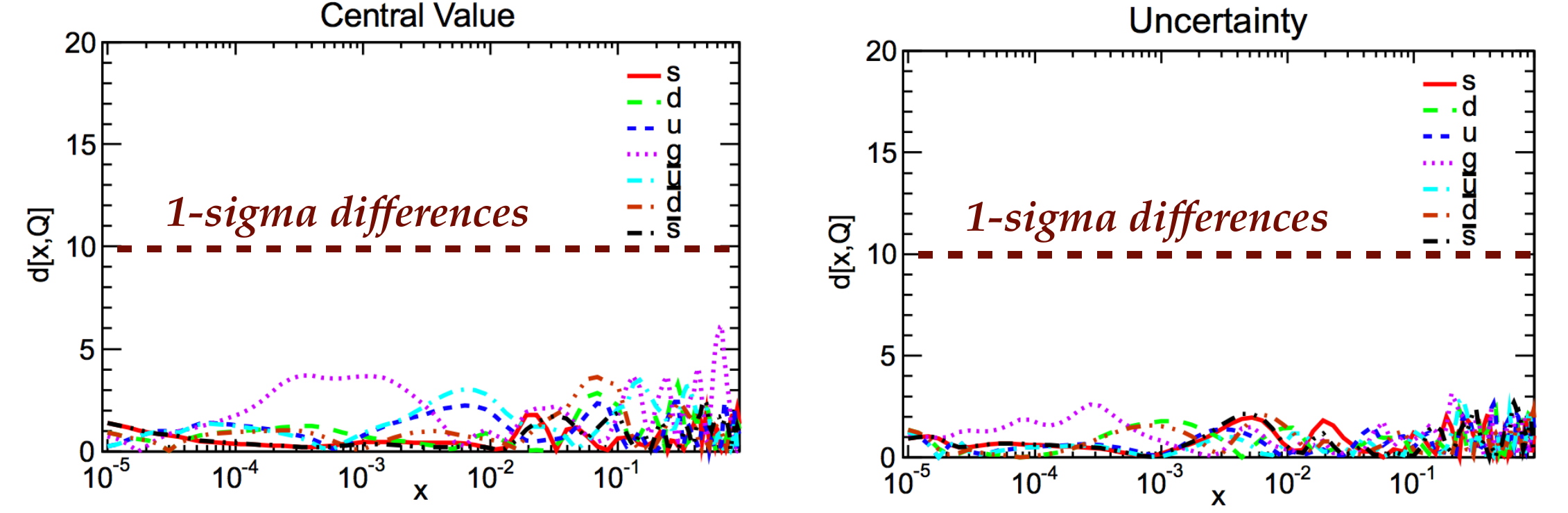}
  \caption{\small The statistical distances between the seven
    PDFs (for both central values and uncertainties)
    in NNPDF3.0 fits using either the 2-5-3-1 or the 2-20-5-1 architecture.
    The dotted line indicates the one-sigma difference for
    the case of $N_{\rm rep}=100$ replicas relevant here.
  }
\label{fig:distances}
\end{figure}

This last property is illustrated in Fig.~\ref{fig:distances}, where
we show the statistical distances~\cite{Ball:2010de} between the seven
    fitted PDFs, for both central values and uncertainties,
    in NNPDF3.0 fits using either the 2-5-3-1 or the 2-20-5-1 architectures.
    The dotted line indicates the value of one-sigma differences for
    the case of $N_{\rm rep}=100$ replicas, relevant in this comparison.
    We thus find that the effect of increasing the number of fitted
    parameters by an order of magnitude is much smaller
    than the PDF uncertainties themselves.

\paragraph{Optimisation and training algorithms.}

The training (also known as learning or optimisation phase) of neural
networks is carried out in most cases using some variant of the gradient
descent method, such as back-propagation~\cite{Forte:2002fg} or stochastic gradient descent.
In these methods, the determination of the fit parameters
(namely the weights and thresholds of the NN) requires the evaluation of the
gradients of $\chi^2$, that is,
\begin{equation}
  \frac{\partial \chi^2}{\partial w_{ij}^{(l)}}
  \,\mbox{,} 
  \quad 
  \frac{\partial \chi^2}{\partial \theta_{i}^{(l)}}
  \,\mbox{.}
\label{eq:param_gradients}
\end{equation}
Computing these gradients in the NNPDF case would be quite involved
due to the non-linear relation between  the fitted experimental data and the
input PDFs, which proceeds through convolutions both with the DGLAP
evolution kernels and the hard-scattering partonic cross-sections
as encoded into the optimised
{\tt APFELgrid} fast interpolation strategy~\cite{Bertone:2013vaa,Bertone:2016lga}.

The theory prediction for a collider cross-section in terms of the NN
parameters reads
 \be
 \label{eq:SMEFTxsec}
 \sigma^{\rm \small (th)}\lp \{ \omega,\theta\}\rp = \widehat{\sigma}_{ij}(Q^2)\otimes \Gamma_{ij,kl}
 (Q^2,Q_0^2) \otimes q_k\lp Q_0,\{ \omega,\theta\} \rp \otimes q_l \lp Q_0 ,\{ \omega,\theta\}\rp
 \ee
 where $\otimes$ indicates a convolution over $x$, $\widehat{\sigma}_{ij}$
 and $\Gamma_{ij,kl}$ stand for the hard-scattering cross-sections
 and the DGLAP evolution kernels respectively, and sum over repeated
 flavour indices is understood.
 In the {\tt APFELgrid}  approach, this cross-section can be expressed in
 a much compact way as
  \be
  \label{eq:thpredSM}
  \sigma^{\rm \small (th)}\lp \{ \omega,\theta\}\rp  = \sum_{i,j=1}^{n_f}\sum_{a,b=1}^{n_x}{\tt FK}_{k,ij,ab} \cdot
  q_i\lp x_a,Q_0, \{ \omega,\theta\}\rp \cdot q_j\lp x_b,Q_0, \{ \omega,\theta\}\rp \,,
  \ee
  where now all perturbative information is pre-computed
  and stored in the ${\tt FK}_{k,ij,ab}$ interpolation tables,
  and $a,b$ run over a grid in $x$.
  The convoluted relation between $\sigma^{(\rm th)}$ and the NN parameters
  in Eq.~(\ref{eq:thpredSM}) is what makes the implementation of gradient
  descent methods challenging.

In the proton NNPDF global analysis, both in the polarised
and the unpolarised case, the NN training is carried out instead by means
 Genetic Algorithms (GAs).
 GAs~\cite{tau} are based on a combination of deterministic and stochastic ingredients
 which make them particularly useful to explore complex parameter spaced
 without getting stuck in local minima, and which do not require the knowledge
 of the $\chi^2$ gradients in Eq.~(\ref{eq:param_gradients}) but
 only of its local values.
 
\begin{figure}[t]
  \centering
  \includegraphics[width=.41\linewidth]{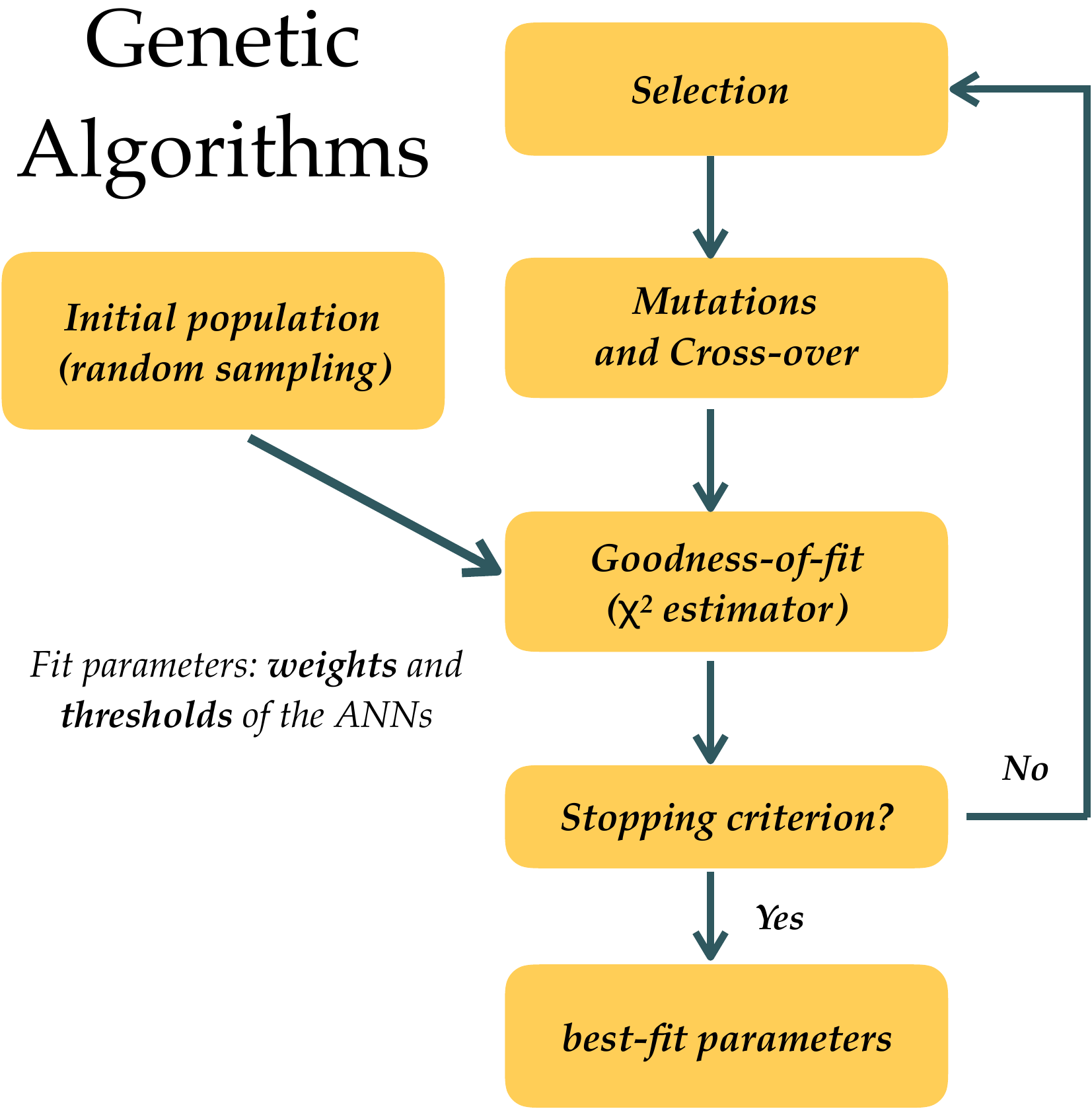}
  \includegraphics[width=.58\linewidth]{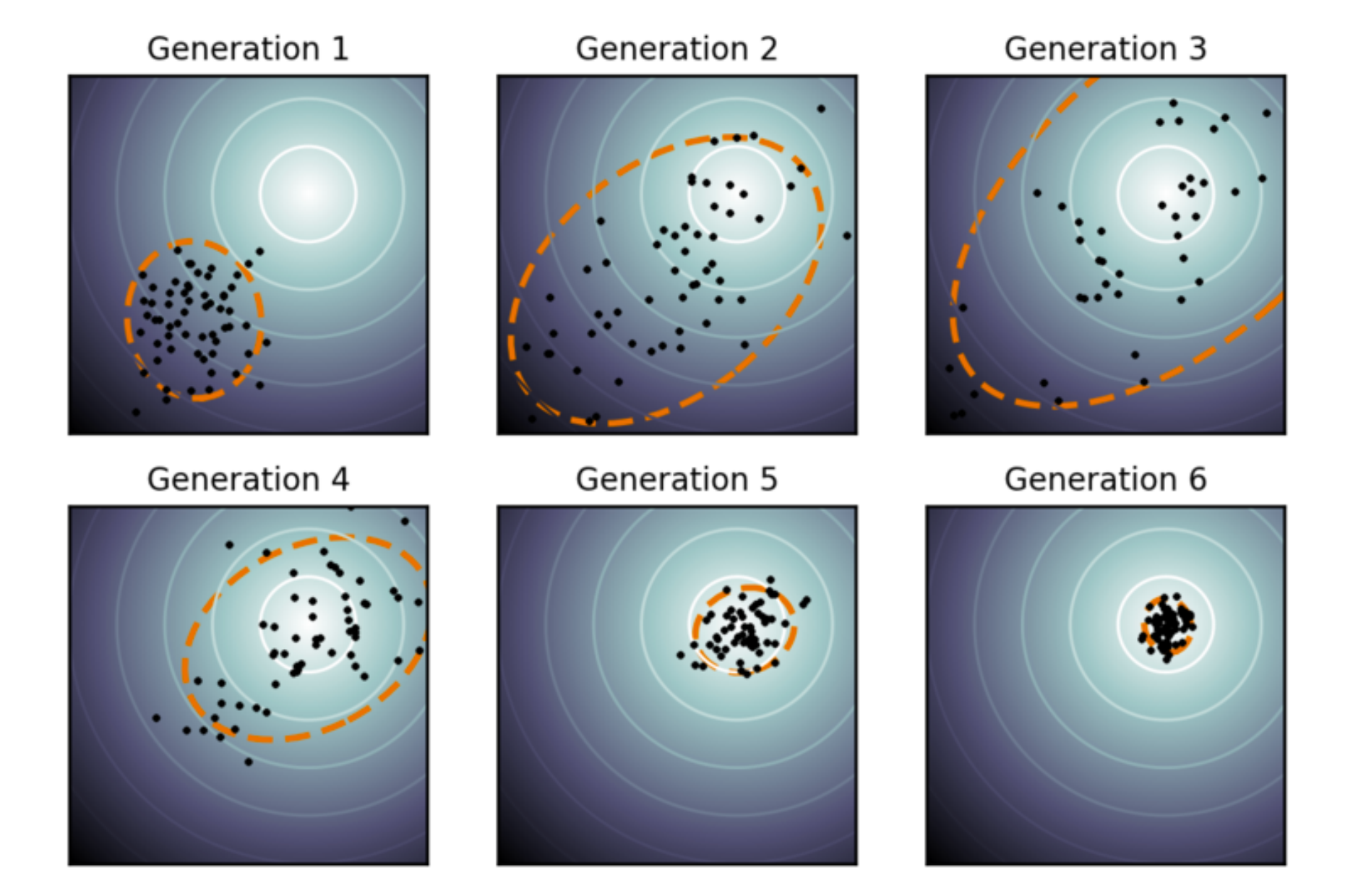}
  \caption{\small Left: Genetic Algorithms work by means
    of an efficient combination of deterministic and stochastic
    ingredients.
    Right: schematic representation of how the CMA-ES algorithm works
    in a toy scenario, showing
    how it manages to approach the global minimum while at the same time
    stochastically sampling the region around it.
  }
\label{fig:CMA-ES}
\end{figure}

As illustrated in Fig.~\ref{fig:CMA-ES},
at each iteration of the fit, variations of the PDF parameters denoted
as PDFs are generated 
by random adjustment of the previous best-fit neural-network parameters.
The mutant PDF parameters with the lowest values of
the figure of merit $\chi^2$ quantifying the agreement with data are then
selected as the best-fit for the next iteration.
Clearly, such procedure is not sensitive to the higher-order structure of the problem
in the parameter space, and thus while being more flexible
and requiring only local $\chi^2$ values, it is bound to be 
less efficient than gradient descent.
In the NNPDF3 family of global analysis, an optimised
GA has been adopted with parameters tuned by means of the closure
tests discussed below. 

Alternative optimisation strategies have been explored within
our Collaboration.
One important example is the  Covariance Matrix Adaption - Evolutionary Strategy 
(CMA-ES) family of algorithms~\cite{Hansen2006,DBLP:journals/corr/Hansen16a},
used in the NNPDF determinations of fragmentation functions.
Here we briefly summarize the main features of this training algorithm.
We denote the set of fit parameters
$\left\{\omega_{ij}^{(l)},\theta_i^{(l)}\right\}$ (weights and thresholds
of the NNs) as a single vector
$\mathbf{a}^{(i)}$.
In the following, the superscript $i$ indicates 
the values at the $i^{th}$ iteration.
The fit parameters are initialised at the beginning of the fit according
to a multi-Gaussian distribution $\mathcal{N}$ with zero mean and unit
covariance
\begin{equation}
  \mathbf{a}^{(0)} \sim \mathcal{N}(0,\mathbf{C}^{(0)})
  \,\mbox{,} 
  \quad 
  \mathbf{C}^{(0)} = \mathbf{I}
  \,\mbox{.}
\label{eq:avect}
\end{equation}
where we use $\sim$ to denote the distribution of the random vector.
This vector is used as the centre of a search distribution in
the fitting parameter space.
At every iteration of the algorithm, $\lambda$ mutants
$\mathbf{x}_1, \ldots, \mathbf{x}_\lambda$ of the NN parameters are generated
by means of the following rule:
\begin{equation}
  \mathbf{x}_k^{(i)} 
  \sim \mathbf{a}^{(i-1)} + \sigma^{(i-1)}\mathcal{N}(0,\mathbf{C}^{(i-1)})
  \,\mbox{,} 
  \qquad 
   k = 1\,\mbox{,} \ldots \mbox{,} \lambda  \, ,
\label{eq:mutgen}
\end{equation}
in other words, mutants are generated around the search centre according to a
multi-Gaussian $\mathcal{N}$ with covariance $\mathbf{C}^{(i)}$ and
according to a step-size $\sigma^{(i)}$.
The mutants are then sorted according to their value
of the figure of merit such that
$\chi^2(\mathbf{x}_k) < \chi^2(\mathbf{x}_{k+1})$.
Subsequently,
the new search
centre is computed as a weighted average over a fixed fraction of the best
mutants
\begin{equation}
  \mathbf{a}^{(i)} 
  = 
  \mathbf{a}^{(i-1)} + \sum_{k=1}^{\mu=\lambda/2} W_k 
  \left( \mathbf{x}_k^{(i)} - \mathbf{a}^{(i-1)} \right)
  \,\mbox{,}
\label{eq:bestmut}
\end{equation}
where the $\{W_k\}$ are parameters of the CMA-ES algorithm which
have been independently tuned for similar applications.

The most important feature of the CMA-ES algorithms is that both the step
size $\sigma^{(i)}$ and the search distribution covariance matrix
$\mathbf{C}^{(i)}$ are being optimised by the fit procedure.
To achieve this, the information present in the ensemble of mutants 
is used to learn preferred directions in parameter space without the need for 
the explicit computation of the gradients of the $\chi^2$.
This adaptive behaviour improves the efficiency of the minimisation procedure 
in comparison to the traditional genetic algorithm described above.
Note also that each iteration's best fit is computed from the
weighted average over a subset of mutants, rather than taking
only the mutant with smaller value of the $\chi^2$.
In this way,  the effects of the possible
statistical fluctuations in the $\chi^2$ are reduced.

In Fig.~\ref{fig:CMA-ES} we display
an schematic representation of how the CMA-ES algorithm works
    in a toy scenario, showing
    how it manages to approach the global minimum while at the same time
    stochastically sampling the region around it.
    Starting from a random population of solutions far from
    the minimum (white region), the spread (variance) of the population
    increases while at the same time the average (center) solution moves closer to
    the minimum.
    As the number of generations increases, the average solution
    remains close to the minimum but now the variance has been reduced
    significantly, indicating that the algorithm has converged.

    As mentioned above, a common feature of both GA and CMA-ES
    is that they do
    not require knowledge (analytical or numerical)
    of the gradients of the $\chi^2$ function in the fitting parameter space.
    There is now ongoing work within the Collaboration towards
    the analytical evaluation of these gradients in terms
    of the NN parameters and the perturbative
    information encoded in Eq.~(\ref{eq:thpredSM}), which would make possible
    adopting 
efficient optimisation techniques such as stochastic gradient
descent and back-propagation.
If successful, the program of adopting gradient-based methods should lead to
a significant speed-up of the NNPDF fits, which in the proton case
now take between one and three days per Monte Carlo replica,
depending on the processor details.

\begin{figure}[t]
  \centering
  \includegraphics[width=.99\linewidth]{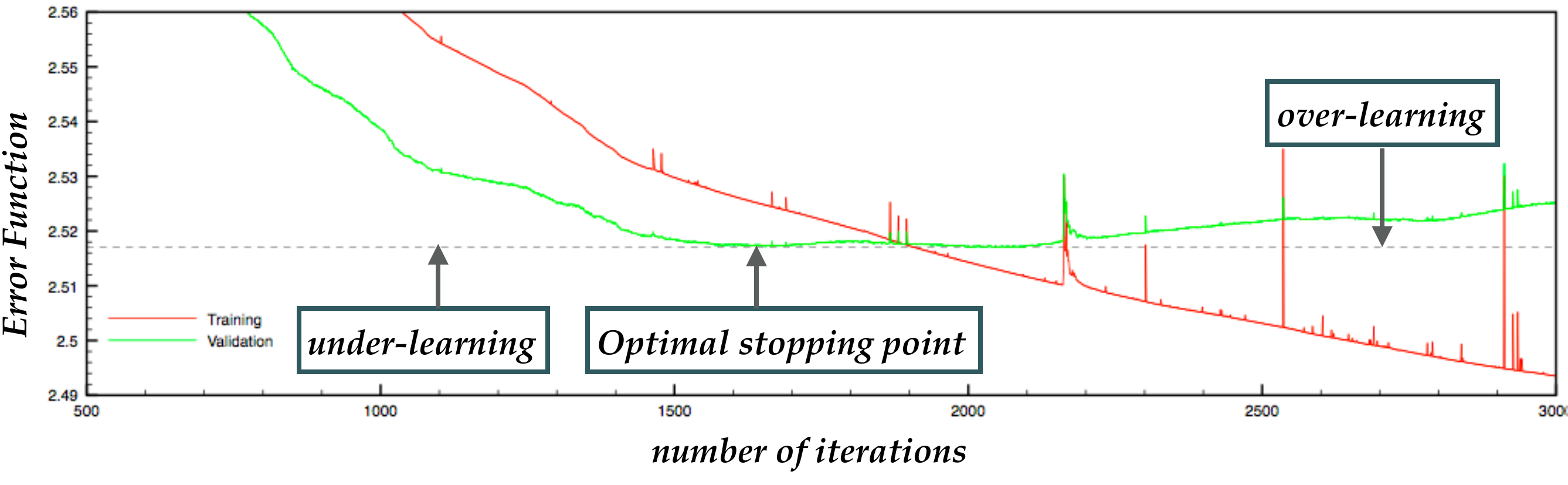}
  \caption{\small Schematic representation of the look-back
    cross-validation stopping used in the NNPDF fits.
  }
\label{fig:stopping}
\end{figure}

The use of a highly redundant parametrisation such
as that illustrated in Fig.~\ref{fig:ANN} raises the worry
that one might end up fitting point-to-point fluctuations.
In order to avoid this situation, known as overfitting  (learning the statistical
fluctuations of the input experimental data rather than the
underlying experimental data), a suitable regularisation strategy
must be adopted.
In the NNPDF fits, we use the look-back cross-validation
stopping criterion, described in~\cite{Ball:2014uwa}, and illustrated
schematically in  Fig.~\ref{fig:stopping}.

This cross-validation stopping strategy works as follows.
First of all, the input experimental measurements are
divided into two categories at random, the training sample
and the validation sample, typically with equal probability.
Only the former is used in the fit, while the latter plays the role
of a control sample used to monitor and validate the training progress.
The optimal stopping point is defined as the global minimum
of the $\chi^2$ of the validation sample, computed over
a large fixed number of iterations (hence the name ``look-back'').
As shown in Fig.~\ref{fig:stopping}, a shorter fit would
result in under-learning (where the NN has not properly learned
yet the underlying law) while a longer fit instead leads to
over-learning (where the NN ends up fitting statistical fluctuations).
The tell-tale sign of the latter is the increase of the validation
$\chi^2$ increases (rather than the decrease) as the number
of iterations increases, indicating that what is being learned
in the training sample is not present in the validation one
(namely the fluctuations).

\paragraph{Closure testing.}
Fitting experimental data is often complicated by a number of factors
unrelated to the methodology itself, such as possible dataset inconsistencies
(either internal or external) or inadequacies of the theoretical
description adopted.
Therefore, it is far from optimal to assess the benefits of an specific
fitting methodology by applying it to the actual
data, while it is much more robust
to test it instead in an analysis of pseudo-data generated from
a fixed (known) underlying theory.

\begin{figure}[t]
  \centering
  \includegraphics[width=.85\linewidth]{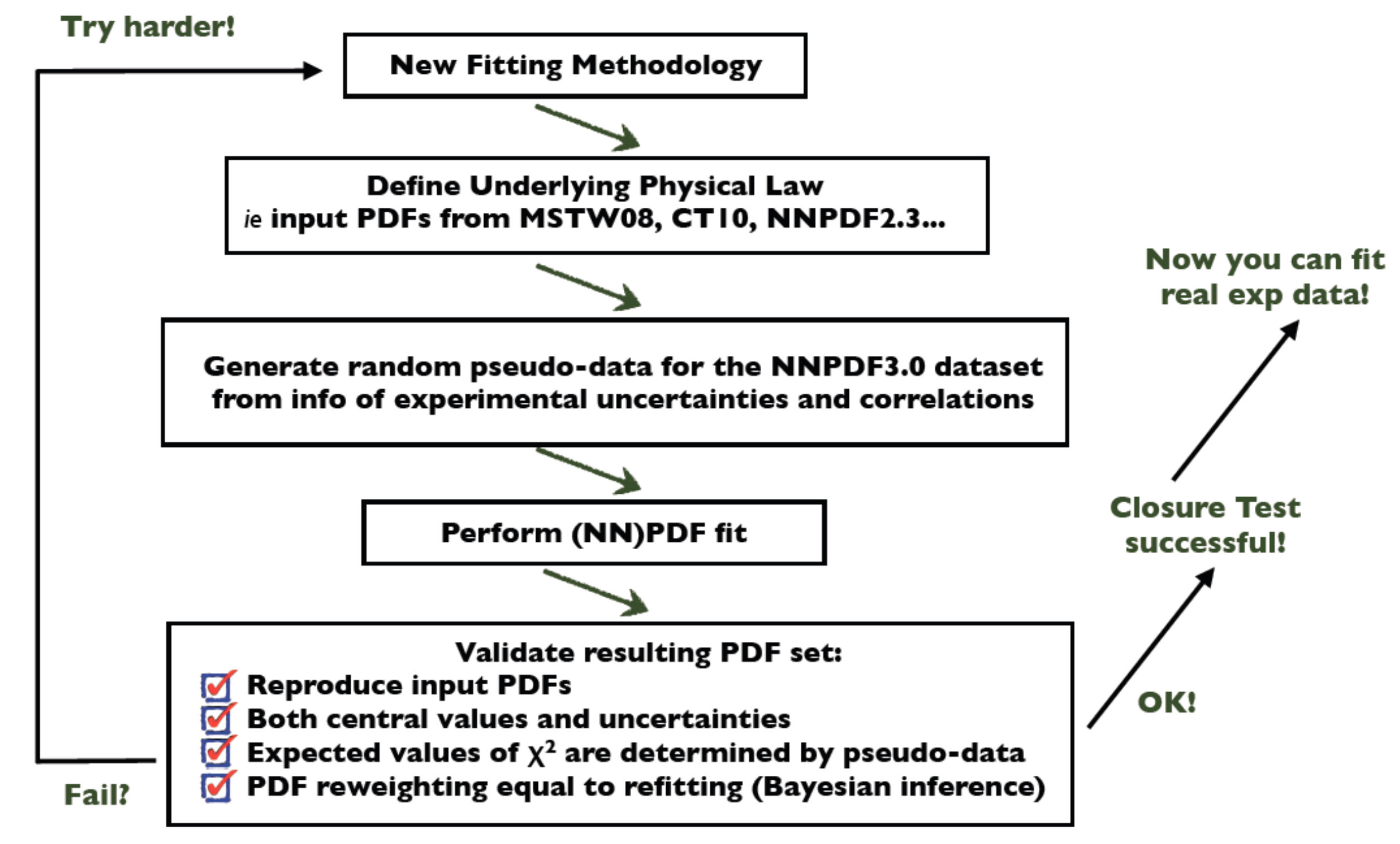}
  \caption{\small Flow-chart indicating how a given PDF fitting
    methodology can be closure tested in order to demonstrate
    the statistical robustness of its results.
  }
\label{fig:CT}
\end{figure}

In these so-called closure tests, one assumes that PDFs at the
input scale $Q_0$
correspond to a specific model (say MMHT14 or CT14), generate
pseudo-data accordingly, and then carry out the NNPDF global fit.
Since in this case the ``true result'' of the fit is known by construction,
it is possible to systematically validate the results verifying
for example that central values are reproduced and fluctuate
as indicated by the PDF uncertainties, or that the $\chi^2$ values
obtained are those that correspond to the generated pseudo-data.
Additionally, one can verify that PDF reweighting based on Bayesian
inference~\cite{Ball:2010gb,Ball:2011gg}
reproduces the fit results, providing a further cross-check
that the resulting PDF uncertainties admit a robust statistical
interpretation.
In Fig.~\ref{fig:CT} we show a flow chart indicating how a given fitted
    methodology can be closure tested in order to demonstrate
    the statistical robustness of its results.

\begin{figure}[t]
  \centering
  \includegraphics[width=.99\linewidth]{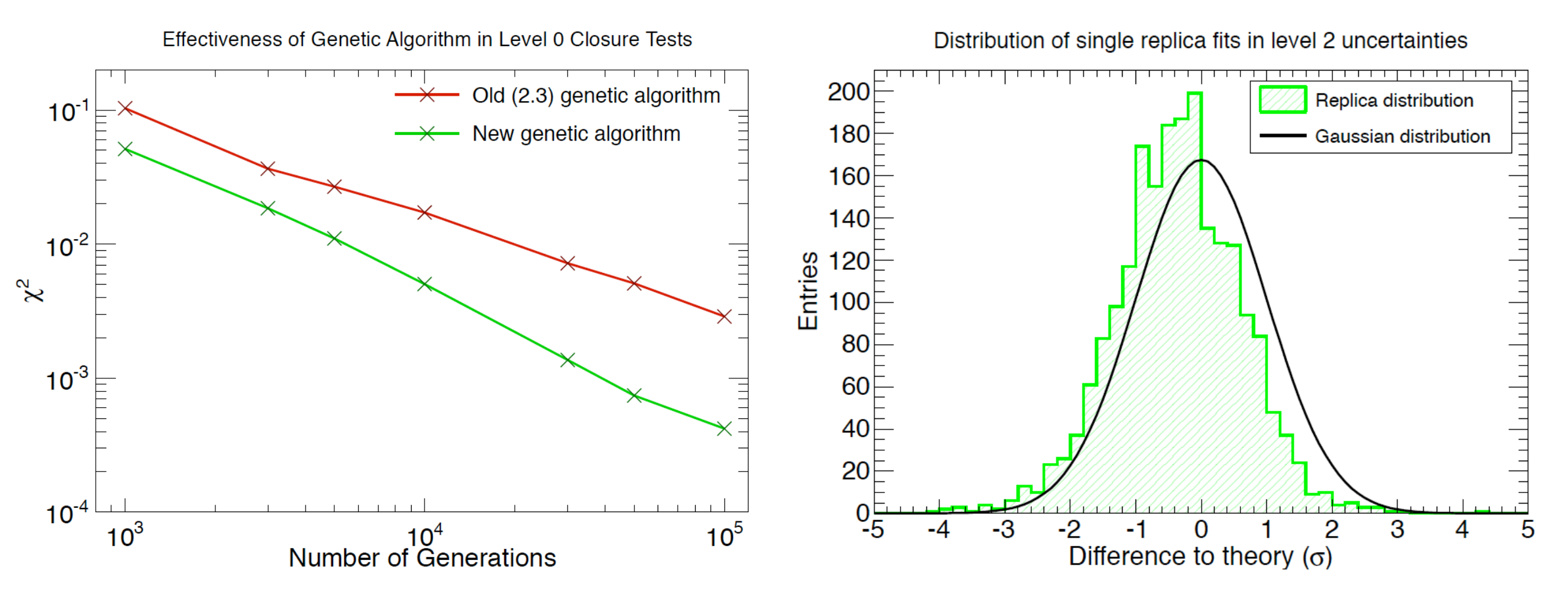}
  \caption{\small Representative results of the closure testing
    of the NNPDF3.0 fits.
    Left: at Level 0, the $\chi^2$ becomes arbitrarily small
    since there exist at least one solution (the input PDF set)
    that corresponds to $\chi^2=0$.
    We also show the improved performance of the NNPDF3.0 GAs as
    compared to the previous one used in the NNPDF2.3 fit.
    Right: the distribution of single replica fits in a Level 2
    closure test, showing that central values fluctuate in accordance
    with the quoted PDF uncertainties as they should.
  }
\label{fig:CT2}
\end{figure}

In Fig.~\ref{fig:CT2} we display some
representative results of the closure testing
    of the NNPDF3.0 fits.
    Closure tests can be performed at Level 0, 1, or 2, depending
    on the amount of fluctuation added on top of the generated
    pseudo-data.
    In particular, in Level 0 closure tests no fluctuations are
    added, while Level 2 corresponds to the same amount of fluctuations
    as in the real global fits (Level 1 is the intermediate case).
    First of all, in the left plot
    we show how at Level 0 the $\chi^2$ becomes arbitrarily small
    since there exist at least one solution (the input PDF set)
    that corresponds to $\chi^2=0$.
    We also show the improved performance of the NNPDF3.0 GAs as
    compared to the previous one used in the NNPDF2.3 fit.
    Secondly, in the right
    plot we display the distribution of single replica fits in a Level 2
    closure test, showing that central values fluctuate in accordance
    with the quoted PDF uncertainties as they should (indicated by the
    good agreement with a Gaussian distribution).
    Several other estimators can be studied to further validate
    the results of these closure test exercises.

\vspace{0.3cm}
\noindent
{\bf Acknowledgments.}
We are grateful to T.~Dorigo and S.~Gleyzer for their invitation to given
this talk at QCHS13.
This work has been supported
by the ERC Starting Grant ``PDF4BSM'' and by the Netherlands
Organisation for Scientific Research (NWO).

\providecommand{\href}[2]{#2}\begingroup\raggedright\endgroup

\end{document}